\documentstyle[psfig]{mn}

\begin{document}
\title[Cluster mass estimates]
{A combined analysis of cluster mass estimates from strong lensing,
X-ray measurement and the universal density profile}
\author[Wu]{Xiang-Ping Wu \\
Beijing Astronomical Observatory and National Astronomical Observatories, 
Chinese Academy of Sciences, Beijing 100012, China\\
}
\date{Accepted 2000 February 22. Received 1999 December 2; in original form 1999 May 4}
\maketitle
\begin{abstract}
We present a combined analysis of mass estimates in the central cores
of galaxy clusters from the strong lensing, the X-ray measurements and
the universal density profile (NFW). 
Special attention is paid to the questions  
(1)whether the previously claimed mass discrepancy 
between the strong lensing and X-ray measurements
is associated with the presence of cooling/non-cooling flows, 
(2)whether the cusped NFW density model can provide a consistent cluster
mass with the strong lensing result and (3)whether a non-zero cosmological 
constant can be of any help to reducing 
the strong lensing --  X-ray mass ratios.
We analyse a sample of 26 arc-like images among 21 clusters,the
X-ray data of which are available in archive.  
The X-ray and NFW cluster masses are obtained by assuming that the intracluster
gas is isothermal and in hydrostatic equilibrium with the underlying 
gravitational potential of the clusters. 

A statistical comparison of these three mass estimates reveals
that the mass discrepancies for all the events are well within 
a factor of $2$, if  X-ray measurement uncertainties are included.
In particular, we confirm the result of Allen that the 
larger mass discrepancy is only detected in the intermediate cooling, 
especially non-cooling flow clusters, 
thus attributing the mass discrepancy to 
the local dynamical activities in the central regions. 
We show that the NFW profile yields a consistent cluster mass
with the conventional X-ray measurement, which is interpreted as 
the consequence of the common working hypothesis behind the two methods. 
Any difference between these two models must occur at even smaller
radii (e.g. within the arc-like images) or at large radii. 
It appears that the introduction of
the cusped density profile as the dark matter distribution of clusters 
cannot raise the cluster masses enclosed within arc-like images.
Finally, a non-zero cosmological constant is able to moderately reduce  
the mass ratios of $m_{\rm lens}$ to $m_{\rm xray}$. These results, 
together with the excellent agreement between 
the X-ray, optical and weak lensing determined cluster masses on scales
greater than the X-ray core sizes found in the early work, indicate
that the mass discrepancy between strong lensing and other methods 
in the intermediate cooling and non-cooling flow clusters  
is likely to have arisen from both the oversimplification of lensing model and 
the inappropriate application of isothermality and 
equilibrium hypothesis in the central regions of clusters where 
the local dynamical activities make a non-negligible contribution
to both mass estimates.
\end{abstract}


\vskip -3in

\section{Introduction}

An accurate estimate of the total gravitating masses of clusters 
of galaxies is crucial for determinations of the mass-to-light 
ratios $M/L$ and the baryon fractions $f_{\rm b}$ of clusters, while
the latter play a potentially important role in the `direct' 
measurement of the mean mass density of the Universe, $\Omega_{\rm M}$.
Four different techniques have been available today to determine 
the masses of clusters: the optical measurements of distribution and
velocity dispersion of cluster galaxies,  the X-ray measurements of 
intracluster gas and its temperature,
the gravitationally distorted images of distant galaxies behind clusters, 
and the numerical simulations of formation and evolution of clusters.
Because traditional cluster mass estimators using optical/X-ray 
observations of galaxies/intracluster gas rely strongly upon 
the assumption of hydrostatic equilibrium, the gravitational lensing
serves as a powerful and efficient tool for both estimating cluster
masses and testing the accuracy and reliability of traditional
methods. It turns out that there is good agreement between the 
gravitational lensing, X-ray and optical determined cluster masses
on scales larger than the X-ray core radii,  within which 
the X-ray method is likely to underestimate cluster masses
by a factor of 2--4 (Wu 1994; Miralda-Escud\'e \& Babul 1995; 
Wu \& Fang 1997; Allen 1998; Wu et al. 1998 and references therein). 
While a number of mechanisms have been suggested for   
the reported mass discrepancy  (Loeb \& Mao 1994; 
Miralda-Escud\'e \& Babul 1995; Allen 1998), 
a satisfactory explanation has not yet
been achieved. It is generally believed that such a mass discrepancy
has probably arisen from either the oversimplification of
strong lensing model for the central mass distributions of clusters
(Bartelmann \& Steinmetz 1996) or the inappropriate application of 
hydrostatic equilibrium hypothesis in the central regions of clusters
(Wu 1994; Wu \& Fang 1997).

In this paper we intend to have a close examination of the issue as to 
whether the reported mass discrepancy is associated with the following
three factors: (1)the presence of cooling/non-cooling flows in clusters,
(2)the cusped dark matter profile, i.e. the so-called universal density 
profile (Navarro, Frenk \& White 1995; hereafter NFW), as the mass 
distribution of clusters, and (3)a non-zero cosmological constant 
($\Omega_{\Lambda}$). 

Our reconsideration of the influence of cooling/non-cooling flows 
on the X-ray mass estimate of clusters is motivated by the pioneering
work of Allen (1998), who has demonstrated that there is an excellent agreement
between the strong lensing and X-ray determined cluster masses for
cooling-flow clusters while the mass discrepancy is only detected among the
non-cooling flow ones owning to the significant offsets between the X-ray
and lensing centers.  If this is the case, the presence or absence of 
cooling-flows can be used as an indicator of whether or not clusters have
reached the state of dynamical relaxation. Meanwhile, the mass discrepancy
between the strong lensing and X-ray measurements can be attributed to the 
local dynamical activities of clusters where the hydrostatic equilibrium 
assumption becomes invalid. However, such a scenario has been argued  
recently by Lewis et al. (1999) based on an analysis of 14 CNOC clusters. They
have shown that the systematic effects owning to cooling-flows, 
non-equilibrium systems and temperature gradients on the average ratio of 
the `dynamical' cluster masses to the X-ray masses do not 
exceed $15$--$20$ per cent. Therefore, further work is needed to clarify 
the issue.

The original purpose of this investigation was to study whether the cusped
NFW profile as the dark matter distribution of clusters can resolve the
discrepancy between the strong lensing and X-ray  mass estimates.
First, the central singularity in the NFW profile is consistent with 
the early claim, based on the modeling and statistics of strongly
distorted images of background galaxies by massive clusters 
(Hammer 1991; Wu \& Hammer 1993;  Grossman \& Saha 1994), that 
dark matter profiles are sharply peaked towards cluster centers. 
Secondly, except for the small core radius there is a striking similarity 
between the distribution of intracluster gas tracing the dark halo of 
the NFW potential and the conventional $\beta$ model 
(Makino, Sasaki \& Suto 1998). Thirdly, many high-resolution simulations
of structure formation have shown that the NFW profile is independent of
mass, initial density fluctuation or cosmology
(e.g. Cole \& Lacey 1996; Navarro, Frenk \& White 1997; 
Eke, Navarro \& Frenk 1998).  In particular, 
Makino \& Asano (1999) have recently applied the NFW profile to 
three lensing clusters, A2163, A2218 and RXJ1347, 
and compared the NFW profile and strong lensing derived cluster masses.
Their study indicates that 
the steeper cusps in the NFW profile can indeed reduce the mass discrepancy 
between the X-ray and gravitational lensing measurements.
On the other hand, the NFW profile can also recover the observed surface 
number density of cluster galaxies (e.g. Carlberg, Yee \& Ellingson 1997),
from which one can fix the two free parameters in the NFW profile and then
work out the total dynamical masses of the clusters.
Using the 14 CNOC distant clusters, Lewis et al. (1999) found no systematic
bias between the (NFW) dynamical and X-ray methods. Consequently,  
the dynamical mass derived from the cusped NFW profile cannot be reconciled 
with the gravitating mass from strong lensing. This implies that 
either the NFW profile may not be extrapolated to smaller radii ($\sim10$ kpc)
or the equilibrium hypothesis may break down in the central regions of
clusters, provided that the current strong lensing model gives  a 
reliable estimate of the cluster masses enclosed with the arc radii.
Alternatively, the giant arc statistics, based on the NFW profile as 
the underlying gravitational potential and the X-ray measurements of 
the gas distribution, can marginally reproduce the observed number of 
giant arcs  but requires unreasonably high X-ray 
temperatures for some clusters (Molikawa et al. 1999). 
Thus, the detailed modeling of strong lensing including the `irregularity'
in the mass distribution of a cluster is once again advocated. 
Taking these results as a whole, we feel that at least one of the following
working hypotheses should be abandoned: 
(1) a simple estimate of cluster mass from
strong lensing, (2) the hydrostatic equilibrium in the central region of 
a cluster, and (3) the NFW profile as the total mass distribution.
In order to stress the point, more examples will be provided in this paper and 
a systematic comparison among the cluster masses given by the strong lensing,
the NFW profile and the X-ray measurements will be made.

Another question we would like to address is how our estimates of the
strong lensing and X-ray cluster masses are affected by 
a non-zero cosmological constant $\Omega_{\Lambda}$.  
This arises because the lensing mass 
$m_{\rm lens}$ is proportional to $D_{\rm d}D_{\rm s}/D_{\rm ds}$ 
while the X-ray mass
$m_{\rm xray}$ goes as $D_{\rm d}$, where  $D_{\rm d}$, $D_{\rm s}$ and 
$D_{\rm ds}$ are 
the angular diameter distances to the cluster, to the background
galaxy, and from the cluster to the galaxy, respectively. 
As a result, the mass ratio $m_{\rm lens}/m_{\rm xray}$ depends on 
the cosmic density parameter $\Omega_{\rm M}$ and $\Omega_{\Lambda}$ 
through $D_{\rm s}/D_{\rm ds}$. In particular, 
$m_{\rm lens}/m_{\rm xray}$ becomes smaller in an  
$\Omega_{\Lambda}$ dominated 
universe than in an $\Omega_{\rm M}$ dominated one.
This opens a new possibility to reduce the large discrepancy between 
$m_{\rm lens}$ and $m_{\rm xray}$. 
We will compare the lensing and X-ray cluster 
masses in a flat cosmological model with and without 
the cosmological constant. Throughout this paper we assume 
$H_0=50$ km s$^{-1}$ Mpc$^{-1}$.

\section{Cluster mass estimates}

\subsection{Strong gravitational lensing}

Gravitational lensing furnishes a simple yet efficient way to measure 
the projected cluster mass along the line of sight. It is believed that
a simple spherical lensing model provides 
rather a good estimate of the projected cluster mass 
within the position ($r_{\rm arc}$) of arc-like image, which reads 
\begin{equation}
m_{\rm lens}(<r_{\rm arc})=\pi r^2_{\rm arc}\Sigma_{\rm crit},
\end{equation}
where $\Sigma_{\rm crit}=(c^2/4\pi G)(D_{\rm s}/D_{\rm d}D_{\rm ds})$ 
is the critical surface mass density.
The above equation is actually the lensing equation for a cluster lens of 
spherical mass distribution with a negligibly small alignment 
parameter for the distant galaxy with respect to $r_{\rm arc}$.  
We adopt the value $m_{\rm lens}(<r_{\rm arc})$
from the detailed modeling of arclike images  where 
available from literature.

\subsection{X-ray measurements}

Assuming that the diffuse X-ray emitting gas in a cluster is isothermal and 
in hydrostatic equilibrium with the underlying gravitational potential 
of the cluster, we have
\begin{equation}
-\frac{GM(r)}{r^2}=\frac{kT}{\mu m_{\rm p}}
       \frac{\rm d \ln n_{\rm gas}(r)}{{\rm d} r},
\end{equation}
where $M(r)$ is the total cluster mass enclosed within radius $r$, 
$T$ and $n_{\rm gas}(r)$ are the gas temperature and number density,
respectively, and $\mu=0.585 $ denotes the mean molecular weight.
If the spatial distribution of intracluster gas is described by 
the conventional
$\beta$ model, $n_{\rm gas}(r)=n_{\rm gas}(0)
         (1+r^2/r_{\rm c}^2)^{-3\beta/2}$
where $r_{\rm c}$ is the core radius of the X-ray gas profile,
we can easily get the dynamical mass distribution from equation (2).
In order to compare with the gravitational lensing result, we 
use the projected X-ray cluster mass within radius $r$ (Wu 1994)
\begin{equation}
m_{\rm {xray}}=1.13\times10^{13}\beta_{\rm {fit}}\tilde{m}(r)
            \left(\frac{r_{\rm c}}{0.1\;{\rm Mpc}}\right)
            \left(\frac{kT}{1\;{\rm keV}}\right)\; {\rm M}_{\odot},
\end{equation}
where 
\begin{eqnarray*}
\tilde{m}(r)=& \frac{(R/r_{\rm c})^3}{(R/r_{\rm c})^2+1} 
               \;\;\;\;\;\;\;\;\;\;\;\;\;\;\;\;\; \\
           & -\int_{r/r_{\rm c}}^{R/r_{\rm c}}x\sqrt{x^2-(r/r_{\rm c})^2}   
             \frac{3+x^2}{(1+x^2)^2}{\rm d}x,
\end{eqnarray*}
and $R$ is the physical radius of the cluster and will be taken to be
$R=3$ Mpc in the actual computation. Our conclusion is unaffected 
by this choice.

\subsection{The universal density profile}

The virialized dark matter halo follows (NFW) 
\begin{equation}
\rho=\frac{\rho_{\rm s}}{(r/r_{\rm s})(1+r/r_{\rm s})^{2}},
\end{equation}
where $\rho_{\rm s}$ and $r_{\rm s}$ are the characteristic density and length,
respectively. 
The projected cluster mass within radius $r$ from equation (4) is simply
\begin{eqnarray}
m_{\rm uni}= & 4\pi \rho_{\rm s} r_{\rm s}^3 \; \cdot\; 
                                                \nonumber \\ 
             &  \left\{
   \begin{array}{ll}
	\ln \frac{r}{2r_{\rm s}} + \frac{r_{\rm s}}{\sqrt{r_{\rm s}^2-r^2}}
                           \ln\frac{r_{\rm s}+\sqrt{r_{\rm s}^2-r^2}}{r}, 
                             & r<r_{\rm s};\\
	\ln \frac{r}{2r_{\rm s}} + \frac{r_{\rm s}}{\sqrt{r^2-r_{\rm s}^2}}
                            \arctan \frac{\sqrt{r^2-r_{\rm s}^2}}{r_{\rm s}}, 
                             & r>r_{\rm s}.
    \end{array} \right. 
\end{eqnarray}
In principle, the two parameters $\rho_{\rm s}$ and $r_{\rm s}$ 
can be determined from  
the observed X-ray surface brightness of intracluster gas 
or the surface number density of galaxies, in combination with the
measurement of X-ray temperature or velocity dispersion of galaxies.
We will work with the X-ray data in this paper.

Inserting the total mass $M(r)$ deduced from the NFW profile into equation (2)
will result in an analytic form of gas profile (Makino et al. 1998):
\begin{equation}
n_{\rm gas}(x)=n_{\rm gas}(0){\rm e}^{-\alpha}(1+x)^{\alpha/x},
\end{equation}
in which $x=r/r_{\rm s}$ and $\alpha=4\pi G \mu m_{\rm p} \rho_{\rm s} 
	r_{\rm s}^2/kT$. 
Since $n_{\rm gas}(\infty)=n_{\rm gas}(0){\rm e}^{-\alpha}$, we introduce a
background subtracted gas number density 
$\tilde{n}(x)=n_{\rm gas}(x)-n_{\rm gas}(\infty)$, which reads
\begin{equation}
\tilde{n}(x)=\frac{\tilde{n}(0)}{{\rm e}^{\alpha}-1}
             \left[(1+x)^{\alpha/x}-1\right].
\end{equation}
This avoids the divergence in the computation of 
X-ray surface brightness below. Recall that an arbitrary cut-off
radius was used by Makino \& Asano (1999) for the same purpose.
The X-ray surface brightness in the scenario of the optically 
thin and isothermal plasma emission is 
\begin{equation}
S_{\rm x}(\theta)\propto \int_{D_{\rm d}\theta/r_{\rm s}}^{\infty}
                   \frac{\tilde{n}^2(x)x{\rm d}x}
                        {\sqrt{x^2-D_{\rm d}\theta/r_{\rm s}}},
\end{equation}
A straightforward computation yields
\begin{eqnarray}
S_{\rm x}(\theta) \propto & \int_{\theta/\theta_{\rm s}}^{\infty}\;\;
                   \frac{\sqrt{x^2-(\theta/\theta_{\rm s})^2}}{x}
                   \;\;(1+x)^{\alpha/x}\;\; \cdot \nonumber \\
	           & \left[(1+x)^{\alpha/x}-1\right]
		    \left[\frac{1}{1+x}-\frac{\ln(1+x)}{x}\right]{\rm d}x,
\end{eqnarray}
in which $\theta_{\rm s}=r_{\rm s}/D_{\rm d}$. 
Our task is thus reduced to finding out the 
two parameters $\alpha$ and $r_{\rm s}$ from the observed X-ray surface 
brightness $S_{\rm x}$ and gas temperatures $T$ of clusters, and then to 
obtain the projected cluster masses from equation (5).

\section{Cluster sample and results}

21 clusters (Table 1) are selected  from the Strong Lensing Cluster Sample
of Wu et al. (1998) by requiring that both the X-ray temperature and
the surface brightness profiles are well measured with $ROSAT$ and/or $ASCA$. 
We have not included those clusters whose temperatures
are estimated by indirect methods such as the X-ray luminosity-temperature
correlation and the velocity dispersion-temperature correlation. 
These 21 clusters have a mean redshift of $\langle z_{\rm d}\rangle=0.30$, and
contain 26 arc-like images. For the 11 arcs that have no redshift 
information, we estimate their lensing masses $m_{\rm lens}$ by assuming
the mean redshifts of $\langle z_{\rm s}\rangle=0.8$ and $2.0$, respectively.
We calculate the values of $m_{\rm lens}(<r_{\rm arc})$ for two cosmological
models: (1) $\Omega_{\rm M}=1$ and $\Omega_{\Lambda}=0$ and 
(2) $\Omega_{\rm M}=0.3$ and $\Omega_{\Lambda}=0.7$.
The resultant cluster masses enclosed within the positions of 26 arc-like 
images are listed in Table 1. We have not provided the uncertainties in 
$m_{lens}$ no matter whether they are derived from the detailed modeling 
of arcs or from the simple spherical assumption.
In principle, there will be no uncertainties associated with $m_{\rm lens}$
in the simple spherical lensing model if redshifts of the lensing clusters
and of the arc-like images are reliably determined. However, 
the presence of substructures and asymmetrical mass distributions 
in  clusters may complicate this simple cluster mass estimate,
which will be the major source of uncertainties in $m_{\rm lens}$.
The absence of the secondary arc-like images in most of the arc-cluster
systems should be a strong argument against the spherical mass
distribution in the central regions of clusters.

\begin{table*}
\vskip 0.2truein
\begin{center}
\caption{Strong Lensing Cluster Sample}
\vskip 0.2truein
\begin{tabular}{ l l l l l l l }
\hline
cluster & $z_{\rm cluster}$ & $z_{\rm arc}$ & 
\multicolumn{2}{c}{$r_{\rm arc}$ (Mpc)} & 
\multicolumn{2}{c}{$m_{\rm lens} (10^{14}{\rm M}_{\odot})$} \\
        &               &           &
\multicolumn{1}{c}{$\Omega_{\rm M}=1$} & 
             \multicolumn{1}{c}{$\Omega_{\rm M}=0.3$} & 
\multicolumn{1}{c}{$\Omega_{\rm M}=1$} & 
            \multicolumn{1}{c}{$\Omega_{\rm M}=0.3$} \\
\hline
      &       &       &       &    &    &    \\
A370$^+$  & 0.373 & 1.3   & 0.35  & 0.41   &  13.0  & 13.6  \\
      &       & 0.724 & 0.16  & 0.19   &  2.90  & 2.99  \\
A963$^+$   & 0.206 & ...   & 0.0517& 0.0567 &  0.25(0.21)$^*$  & 0.25(0.21)$^*$  \\
      &       & 0.711 & 0.080 & 0.088  &  0.60  & 0.61  \\
A1689 & 0.181 & ...   & 0.183 & 0.199  &  3.6(3.0)$^*$   & 3.7(3.2)$^*$   \\
A1835 & 0.252 &  ...  & 0.150 & 0.167  &  1.98(1.54)$^*$  & 2.02(1.60)$^*$ \\
A2163 & 0.203 & 0.728 & 0.0661& 0.0724 &  0.41  & 0.42  \\
A2218$^+$  & 0.171 & 1.034 & 0.26  & 0.28   &  2.70  & 2.75  \\
      &	      & 0.702 & 0.0794& 0.0859 &  0.623 & 0.632 \\
      &       &	2.515 & 0.0848& 0.091  &  0.570 & 0.587 \\
A2219$^+$  & 0.228 & ...   & 0.079 & 0.087  &  0.517(0.415)$^*$ & 0.527(0.429)$^*$\\
      &       & ...   & 0.110 & 0.122  &  1.60(1.28)$^*$  & 1.63(1.32)$^*$  \\
A2390 & 0.228 & 0.913 & 0.177 & 0.196  &  2.54  & 2.60  \\
CL0024& 0.391 & 1.675 & 0.220 & 0.257  &  2.6   & 2.7 \\     
CL0500& 0.327 &  ...  & 0.15  & 0.17   &  1.90(1.33)$^*$  & 1.95(1.39)$^*$  \\
CL2236& 0.552 & 1.116 & 0.0876& 0.107  &  0.30  & 0.32  \\
CL2244& 0.328 & 2.236 & 0.0465& 0.0532 &  0.20  & 0.21  \\
MS0302& 0.423 & ...   & 0.122 & 0.144  &  1.60(0.95)$^*$  & 1.66(1.01)$^*$  \\
MS0440& 0.197 & 0.530 & 0.089 & 0.097  &  0.89  & 0.90  \\
MS0451& 0.539 & ...   & 0.190 & 0.230  &  5.2(2.3)$^*$   & 5.4(2.4)$^*$  \\
MS1008& 0.306 & ...   & 0.26  & 0.30   &  6.1(4.4)$^*$   & 6.3(4.6)$^*$   \\
MS1358& 0.329 & 4.92  & 0.121 & 0.139  &  0.827 & 0.882 \\
MS1455& 0.257 &   ... & 0.098 & 0.110  &  0.86(0.67)$^*$  & 0.88(0.69)$^*$  \\
MS2137& 0.313 & ...   & 0.0874& 0.0995 &  0.71(0.51)$^*$  & 0.73(0.53)$^*$  \\
PKS0745&0.103 & 0.433 & 0.0459& 0.0482 &  0.30  & 0.30  \\  
RXJ1347&0.451 & 0.81  & 0.24  & 0.28   &  4.2   & 4.4  \\
       &      &       &       &  \\      
 \hline                                                         
\multicolumn{7}{l}{$^+$Multiple-arc system.}\\
\multicolumn{7}{l}{$^*$Arc-like image is assumed at 
$z_{\rm s}=0.8$ ($z_{\rm s}=2$).}
\end{tabular}
\end{center}
 \end{table*}

The X-ray surface brightness profiles $S_{\rm x}$ for our 21 clusters 
are available in archive ($ROSAT$ and $ASCA$),
which have been analysed by a number 
of authors for different purposes. Because the accuracy of our
determination of the dynamical cluster mass from X-ray measurement
is closely connected to the issue as to how precisely we can describe 
the observed X-ray surface brightness $S_{\rm x}$, 
two conventional methods are employed in the fitting of $S_{\rm x}$
(e.g. Neumann \& Arnaud 1999):
(1) a $\beta$ model fit to the entire X-ray surface brightness;
(2) a $\beta$ model fit by excising the central region until an improvement of
$\chi^2$ can be achieved. The latter is usually applied to 
the cooling-flow cluster where  a sharp peak is seen in the X-ray emission 
concentrated in the core of the cluster.
In Table 2 we list the best-fitting parameters
$\beta$ and $r_{\rm c}$ from the first fitting method, while 
the results obtained from the second method are presented in Table 3 for 
14 clusters where available in literature. Here we make no
attempt as far as possible to extrapolate the original work.
For all the clusters we take the temperature data from the literature  
(see Wu, Xue \& Fang 1999; and references therein). 
In order to explicitly demonstrate the possible effect of cooling/non-cooling
flows on the mass ratios of $m_{\rm lens}/m_{\rm xray}$, 
the sample is classified 
into the massive cooling-flow (MC) clusters, the intermediate cooling-flow
(IC) clusters and non-cooling flow (NC) clusters 
in terms of their cooling times (see Allen 1998). Roughly speaking, 
the MC, IC and NC clusters in our sample have 
average mass deposition rates of   
$\ga1000{\rm M}_{\odot}$ yr$^{-1}$, $\sim100 {\rm M}_{\odot}$ yr$^{-1}$ and
$\la10 M_{\odot}$ yr$^{-1}$, respectively.
The projected dynamical masses ($m_{\rm xray}$) of clusters within 
the arc positions are computed according to equation (3). 
Finally, the mass ratio of 
$m_{\rm lens}$ to $m_{\rm xray}$ is obtained for each case and listed in
Tables 2 and 3, in which the error bars reflect 
the combined measurement uncertainties of  $T$, $r_{\rm c}$ and $\beta$. 

\bigskip
\begin{table*}
\vskip 0.2truein
\begin{center}
\caption{X-ray properties: a standard $\beta$ fit to 
the entire X-ray surface brightness}
\begin{tabular}{c}
PLEASE PLACE TABLE 2 HERE\\
\end{tabular}
\vskip 0.2truein
\end{center}
 \end{table*}
\bigskip

\bigskip

\begin{table*}
\vskip 0.2truein
\begin{center}
\caption{X-ray properties: cooling flow regions excluded}
\begin{tabular}{c}
PLEASE PLACE TABLE 3 HERE\\
\end{tabular}
\end{center}
 \end{table*}

\bigskip

In principle, we can get the two parameters $\alpha$ (or $\rho_{\rm s}$) 
and $r_{\rm s}$
in the NFW profile by fitting the observed $S_{\rm x}$ to the predicted
form of equation (9) for all the cases.  
However, the actual operation turns out to be difficult 
for some clusters owning to the instrumental PSF of the X-ray telescopes
($ASCA$, $ROSAT$). 
So, for eight clusters in our sample we estimate the parameters $\alpha$ and
$r_{\rm s}$ using the empirical formula (Ettori \& Fabian 1999),
$\alpha=14.34\beta$ and $r_{\rm s}=3.17r_{\rm c}$. For the three clusters 
(A2219, MS1358 and MS2137), we take the best-fitting values from 
Ettori \& Fabian (1999) who have excluded the cooling-flow regions
in their fitting. For the rest clusters, we adopt the results of
Wu \& Xue (2000), which are obtained by the $\chi^2$ fit of the 
entire data points of $S_{\rm x}$ from the 
Mohr, Mathiesen \& Evard (1999) sample 
to the theoretical predictions (equation 9). 
The  parameters, $\alpha$ and $r_{\rm s}$, for all the clusters of our sample 
are listed in Table 4,  together with the derived
cluster masses within the positions of arcs and their ratios to
the corresponding masses from strong lensing for a cosmological 
model of $\Omega_{\rm M}=1$ and $\Omega_{\Lambda}=0$.

\begin{table*}
\vskip 0.2truein
\begin{center}
\caption{Cluster properties: the NFW profile}
\vskip 0.2truein
\begin{tabular}{ l l l l l  }
\hline
\multicolumn{1}{c}{cluster} &
\multicolumn{1}{c}{$\alpha$} &
\multicolumn{1}{c}{$r_{\rm s}$(Mpc)} &
\multicolumn{1}{c}{$m_{\rm NFW}(<r_{\rm arc})^*$}  & 
\multicolumn{1}{c}{$m_{\rm lens}/m_{\rm NFW}$}  \\
\hline
     &       &      &     &   \\
A370 &  $13.62^{+10.76}_{-5.02}$(3) &  $1.52^{+1.20}_{-0.70}$(3) & 
	$2.57^{+19.89}_{-2.27}$     &   $5.05^{+37.20}_{-4.47}$\\
     &                              &                            & 
        $0.77^{+6.58}_{-0.68}$      &  $3.78^{+29.98}_{-3.38}$  \\
A963 &  $7.31^{+0.57}_{-0.57}$(3)   &  $0.31^{+0.07}_{-0.07}$(3) &
        $0.15^{+0.15}_{-0.07}$      &  $1.68^{+1.62}_{-0.85}$   \\
     &                              &                            & 
        $0.29^{+0.28}_{-0.14}$      &  $2.10^{+1.96}_{-1.04}$  \\
A1689 &  $10.93^{+0.44}_{-0.44}$(1)&  $0.72^{+0.07}_{-0.07}$(1)& 
        $1.44^{+0.58}_{-0.40}$      &  $2.51^{+0.98}_{-0.72}$\\
A1835 &  $9.00^{+0.69}_{-0.69}$(1)  &  $0.23^{+0.05}_{-0.05}$(1)& 
 	 $1.44^{+1.51}_{-0.76}$     &  $1.37^{+1.52}_{-0.70}$\\
A2163 &  $8.83^{+0.16}_{-0.16}$(1)  &  $1.03^{+0.06}_{-0.06}$(1)& 
	 $0.31^{+0.08}_{-0.06}$     &  $1.34^{+0.36}_{-0.28}$\\
A2218 &  $9.19^{+0.33}_{-0.33}$(1)  &  $0.88^{+0.08}_{-0.08}$(1)& 
	 $1.44^{+0.49}_{-0.36}$     & $1.87^{+0.63}_{-0.47}$\\
      &                             &                            & 
	 $0.23^{+0.09}_{-0.07}$     & $2.67^{+0.97}_{-0.71}$\\
      &                             &                            & 
         $0.26^{+0.09}_{-0.07}$     & $2.19^{+0.79}_{-0.59}$\\
A2219 &  $11.51$(2)                 & $1.59$(2)                 & 
         $0.34^{+0.01}_{-0.01}$     & $1.51^{+0.06}_{-0.06}$\\
      &                             &                            & 
         $0.59^{+0.02}_{-0.02}$     & $2.69^{+0.11}_{-0.10}$\\
A2390 &  $8.76^{+0.23}_{-0.23}$(1)  & $0.54^{+0.03}_{-0.03}$(1) &
	 $1.52^{+0.45}_{-0.36}$     & $1.67^{+0.51}_{-0.38}$\\
CL0024&  $6.81^{+1.08}_{-0.72}$(3)  & $0.21^{+0.12}_{-0.08}$(3)    &
         $1.00^{+4.52}_{-0.80}$     & $2.59^{+9.96}_{-2.12}$ \\
CL0500 & $12.91^{+7.17}_{-4.3}$(3)  & $1.30^{+0.67}_{-0.76}$(3)  &
         $0.73^{+9.16}_{-0.65}$     & $2.61^{+22.01}_{-2.42}$\\
CL2236 & $7.60^{+2.58}_{-1.29}$(3)  & $0.21^{+0.13}_{-0.09}$(3)     &
         $0.40^{+2.00}_{-0.33}$     & $0.75^{+3.39}_{-0.63}$\\
CL2244 & $3.81^{+0.09}_{-0.09}$(3)  & $0.22^{+0.10}_{-0.10}$(3)  &
	 $0.085^{+0.298}_{-0.064}$  & $2.36^{+7.48}_{-1.84}$\\
MS0302 & $7.30^{+1.27}_{-1.27}$(1)  & $0.17^{+0.09}_{-0.09}$(1) &
         $0.46^{+1.60}_{-0.36}$     & $3.51^{+14.04}_{-2.73}$\\
MS0440 & $5.62^{+0.17}_{-0.17}$(1)  & $0.075^{+0.011}_{-0.011}$(1)&
         $0.31^{+0.22}_{-0.12}$     & $2.84^{+1.85}_{-1.18}$\\
MS0451 & $5.32^{+0.56}_{-0.56}$(1)  & $0.44^{+0.14}_{-0.14}$(1) &
         $1.01^{+1.69}_{-0.63}$     & $5.17^{+8.63}_{-3.24}$\\
MS1008 & $9.03^{+1.58}_{-1.00}$(3)  & $0.64^{+0.20}_{-0.15}$(3)  & 
         $1.64^{+3.18}_{-1.03}$     & $3.72^{+6.38}_{-2.46}$\\
MS1358 &  $14.29$(2)                & $1.48$(2)                 & 
         $0.55^{+0.31}_{-0.31}$     & $1.51^{+2.03}_{-0.55}$\\
MS1455 & $7.34^{+0.43}_{-0.43}$(1)  & $0.13^{+0.045}_{-0.045}$(1)& 
         $0.44^{+0.60}_{-0.26}$     & $1.96^{+2.75}_{-1.13}$\\
MS2137 & $11.48$(2)                 & $0.18$(2)                 & 
         $0.45^{+0.04}_{-0.07}$     & $1.59^{+0.31}_{-0.13}$ \\  
PKS0745& $8.93^{+0.10}_{-0.10}$(1)  & $0.27^{+0.01}_{-0.01}$(1) & 
         $0.23^{+0.08}_{-0.06}$     & $1.29^{+0.44}_{-0.35}$ \\
RXJ1347& $8.17^{+0.57}_{-0.20}$(3)  & $0.18^{+0.04}_{-0.04}$(3)  & 
         $2.64^{+2.07}_{-1.10}$     & $1.59^{+1.14}_{-0.70}$\\
 & & & & \\
 \hline        
\multicolumn{5}{l}{Nores. $^*$In units of $10^{14}{\rm M}_{\odot}$.}\\
\multicolumn{5}{l}{(1)From Wu \& Xue (2000).}\\
\multicolumn{5}{l}{(2)From Ettori \& Fabian (1999).}\\
\multicolumn{5}{l}{(3)Estimated from $\alpha$-$\beta$
       and $r_{\rm s}$-$r_{\rm c}$ relations (Ettori \& Fabian 1999).}
\end{tabular}
\end{center}
 \end{table*}

We first compare in Fig.1 the projected cluster masses within 
arc-like images determined from strong lensing and X-ray measurements,
using the data in Table 2, where the values of $m_{\rm lens}$ 
for $z_{\rm s}=2$
are chosen if redshifts of the arcs remain unknown. 
It appears that the ratios of $m_{\rm lens}$ to $m_{\rm xray}$ display 
rather large dispersions for many clusters, which is mainly due 
to the fact that the X-ray surface brightness profiles 
of these clusters (e.g. A370, CL0500, etc.) have not been well constrained.
Yet, the mass ratios of all the data points essentially satisfy  
$m_{\rm lens}/m_{\rm xray}<2$ when their uncertainties are taken into account.
Namely, the lensing and X-ray determined cluster masses  are still
consistent with each other within a factor of $2$,
and there is no large discrepancy between these two mass estimates.
Additionally, all the clusters which exhibit a relatively large 
mass ratio of $m_{\rm lens}/m_{\rm xray}>2$, 
if the error bars are neglected for 
the moment, are the NC or IC clusters, 
in contrast with the MS clusters whose mass ratios 
$m_{\rm lens}/m_{\rm xray}$ are roughly consistent with unity. 
We have thus confirmed the  results of Allen (1998).
Alternatively, the introduction of a non-zero cosmological constant 
does help to reduce, though moderately,
the mass discrepancy between $m_{\rm lens}$ and $m_{\rm xray}$.

Next we demonstrate how our X-ray mass measurements within $r_{\rm arc}$ 
are affected by the two different fittings of the X-ray surface 
brightness profiles:  whether or not the excess X-ray emission in the
central cores is excluded. We display in Fig.2 the projected X-ray masses 
within 17 arc-like images obtained from these two methods for a
cosmological model of $\Omega_{\rm M}=1$ and $\Omega_{\Lambda}=0$.
It is immediate that these two methods provide essentially a consistent 
cluster mass at the central core. Consequently, the mass ratios 
$m_{\rm lens}/m_{\rm xray}$ will remain approximately the 
same as those in Table 2.
even if the goodness of a $\beta$ model fit is improved by omitting 
some data points of $S_{\rm x}$ in the central regions.

Finally, we compare the  projected cluster masses within 
arc-like images derived from the strong lensing, the X-ray measurement 
based on a $\beta$ model and the NFW profile. 
Fig.3 shows the results of $m_{\rm NFW}(<r_{\rm arc})$ 
versus $m_{\rm xray}(<r_{\rm arc})$ 
for our sample of 26 arcs among  21 clusters. 
Regardless of the large error bars, the cluster masses given by the NFW
profile are in good agreement with the values from the X-ray analysis.
Indeed, this result contrasts with our initial yet naive speculation 
that the cusped NFW profile at the cluster center can give rise to 
a very different cluster mass enclosed within the arc-like image.
On the other hand, such a situation is not surprising because 
both the X-ray and NFW cluster mass estimates are based on 
the measurements of X-ray surface brightness profiles of clusters 
and the assumption that the X-ray emitting gas  is isothermal and
in hydrostatic equilibrium with the underlying gravitational potential
of the clusters. Recall that the NFW profile can almost recover the
conventional $\beta$ model for the observed X-ray surface brightness 
profiles of clusters. This explains the good agreement between these 
two mass estimates. The similar result has been reported by 
Lewis et al (1999) if the surface number density of galaxies and
their velocity dispersion, instead of the X-ray surface brightness profile 
of intracluster gas and its temperature, are used to calibrate the
NFW profile. As a result, the cusped NFW profile is of no help
to resolving the mass discrepancy between the strong  lensing and X-ray 
measurements, and the ratios of $m_{\rm lens}/m_{\rm NFW}$ listed in Table 4
remain roughly the same as the values of  $m_{\rm lens}/m_{\rm xray}$ 
in Table 2.
If there are any differences between the NFW and X-ray mass estimates,
they are certainly beyond the regions accessible to current lensing
and X-ray data. It could be misleading if one simply extrapolates
our present mass comparison into smaller (or larger) radii.

\section{Discussion and conclusions}

We have conducted a combined analysis of the cluster mass estimates  
from the strong lensing, the X-ray measurement and the NFW profile,
aiming at reducing the mass discrepancy between
the strong lensing and X-ray measurements reported in literature 
(see Wu \& Fang 1997).
In particular, the cusped NFW profile at the central regions of clusters
was thought to be promising for providing a new insight into the problem.
Basically, these three mass estimates originated from very different 
motivations, and should in principle be independent of each other.
However, the actual application of the NFW profile needs the 
priori calibration of the two free parameters, $\rho_{\rm s}$ and $r_{\rm s}$,
which requires the knowledge about the baryonic mass distributions 
(galaxies and intracluster gas) in clusters, in conjunction 
with the spectroscopic measurements of the velocity dispersion/temperature 
of the clusters. So, when the X-ray data are used to fix the NFW profile, 
there is a common working hypothesis behind the  X-ray and NFW mass estimates:
i.e., the hydrostatic equilibrium and isothermality
for the intracluster gas. Therefore, the gravitational lensing appears to be
the unique mass estimate at present independently of the dynamical
state of clusters.

From a detailed comparison of the projected cluster masses 
enclosed within 26 arc-like images among 21 clusters obtained from
the three mass estimates, we have arrived at the following conclusions:

(i)The mass discrepancy among the three methods are actually within a
factor of $2$, if X-ray measurement uncertainties are included.
This is particularly true if a non-zero cosmological constant is
invoked.

(ii)We have confirmed the finding by Allen (1999) that the mass ratios of 
$m_{\rm lens}$ to $m_{\rm xray}$ (or $m_{\rm NFW}$) are consistent with unity 
for the massive cooling flow clusters,
indicating that they are the more dynamically relaxed systems. 

(iii)Our X-ray mass estimates within $r_{\rm arc}$ are little affected by
whether or not the excess X-ray emission in the central cores is excluded
in the fit of the observed X-ray surface brightness. 

(iv)The cluster masses within the arc-like images determined from
the X-ray and NFW methods show an excellent agreement. 
Any differences between these two methods must occur at even smaller
radii ($<r_{\rm arc}$) or at large radii. This arises because both methods
have assumed the isothermality and hydrostatic equilibrium for the 
intracluster gas. In other words, the employment
of the cusped NFW profile for the dark matter distribution of clusters
cannot resolve the reported mass discrepancy between
the strong lensing and X-ray measurements claimed in literature.

Taking these results as a whole, 
together with the previous finding that there is good agreement between 
the X-ray, optical and weak lensing determined cluster masses on scales
greater than the X-ray core sizes 
(Wu \& Fang 1997; Allen 1998; Wu et al. 1998),
we feel that the reported mass discrepancy in the IC, especially
NC clusters is likely to have arisen from 
the oversimplification of the modeling of central mass distributions of the
clusters as lenses. Indeed, if the presence of cooling flows 
in the central cores of clusters is a reliable indicator of their dynamical 
relaxation (Allen 1998), it would not be difficult to understand the 
reported mass discrepancy in the central regions of the IC and NC clusters 
where the local dynamical activities play 
a dominant role in the organization of the central matter distributions.
Consequently, a simple spherical lens model 
may lead to an overestimate of the gravitating masses of the 
lensing clusters (e.g. Bartelmann \& Steinmetz 1996; Allen 1998; 
Wu et al. 1998; etc.), and meanwhile the X-ray mass estimate 
including the NFW method should also allow the intracluster gas to 
deviate from the isothermal and hydrostatic equilibrium distributions.


\section*{Acknowledgments}

I would like to thank Simon White for pointing to me this research, and
an anonymous referee for valuable comments and suggestions. 
This work was supported by 
the National Science Foundation of China, under Grant No. 1972531.

\newpage

\begin{figure*}
\centerline{\hspace{3cm}\psfig{figure=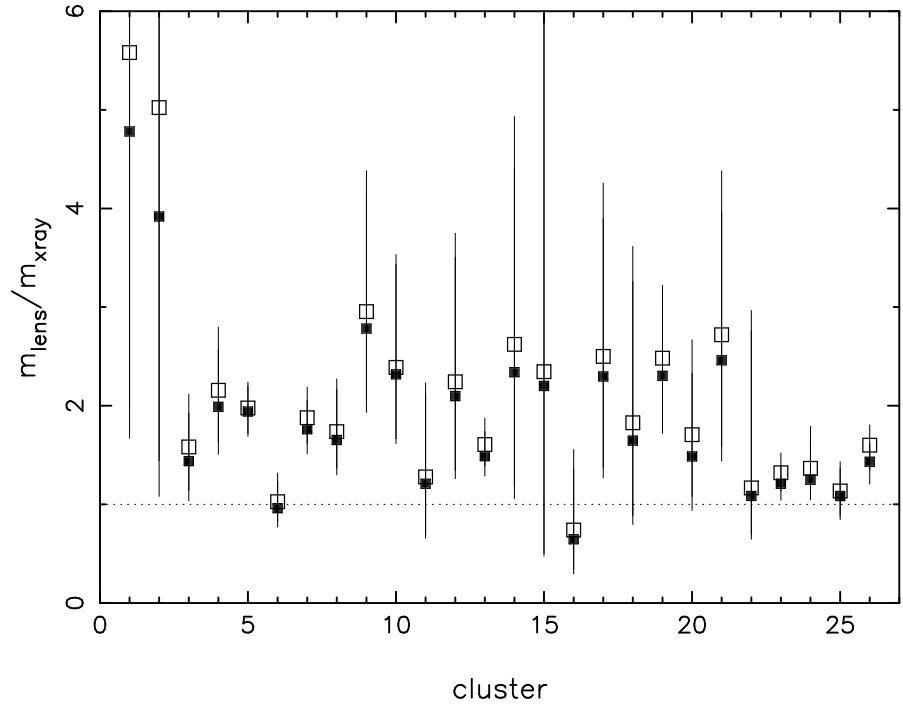,width=1.\textwidth,angle=270}}
\caption{Ratios of the strong lensing and X-ray determined cluster masses
enclosed within 26 arc-like images among 21 clusters.
The horizontal axis is ordered in terms of the arc list in Table 1. 
A flat cosmological model with ($\Omega_{\Lambda}=0.7$, filled squares)
and without ($\Omega_{\Lambda}=0$, open squares) the cosmological
constant is used. For the 11 arc-like images whose redshifts are
not available, we adopt a mean value of $z_{\rm s}=2$.
}
\end{figure*}

\begin{figure*}
\centerline{\hspace{3cm}\psfig{figure=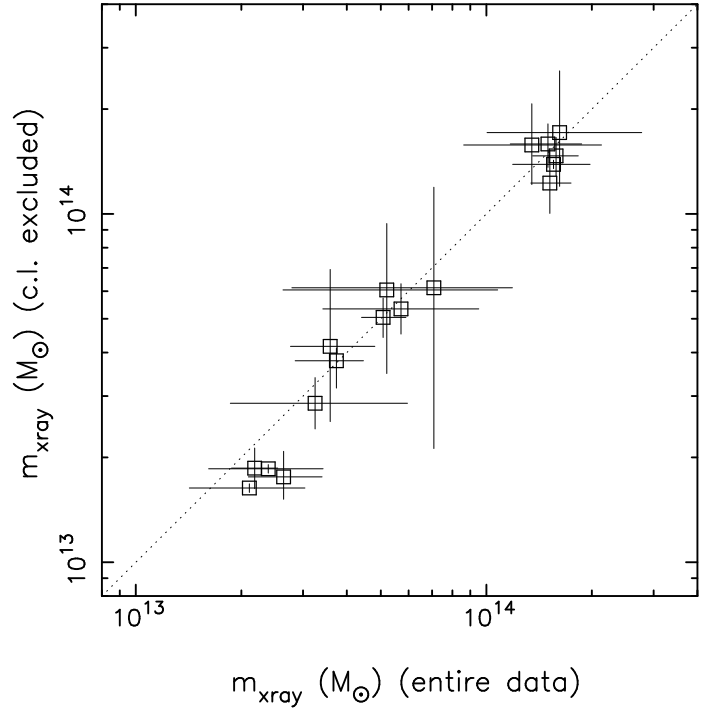,width=1.0\textwidth,angle=270}}
\caption{A comparison of the X-ray cluster masses enclosed within 
17 arc-like images among 14 clusters (Table 3) derived from two 
different fittings of the X-ray surface brightness $S_{\rm x}$ of the clusters:
(1) a $\beta$ model fitted to the entire  $S_{\rm x}$ (horizontal axis), and 
(2) a $\beta$ model fit by excising the central region of $S_{\rm x}$ until 
an acceptable $\chi^2$ is achieved (vertical axis).   
A flat cosmological model of $\Omega_{\rm M}=1$ is assumed.
}
\end{figure*}

\begin{figure*}
\centerline{\hspace{3cm}\psfig{figure=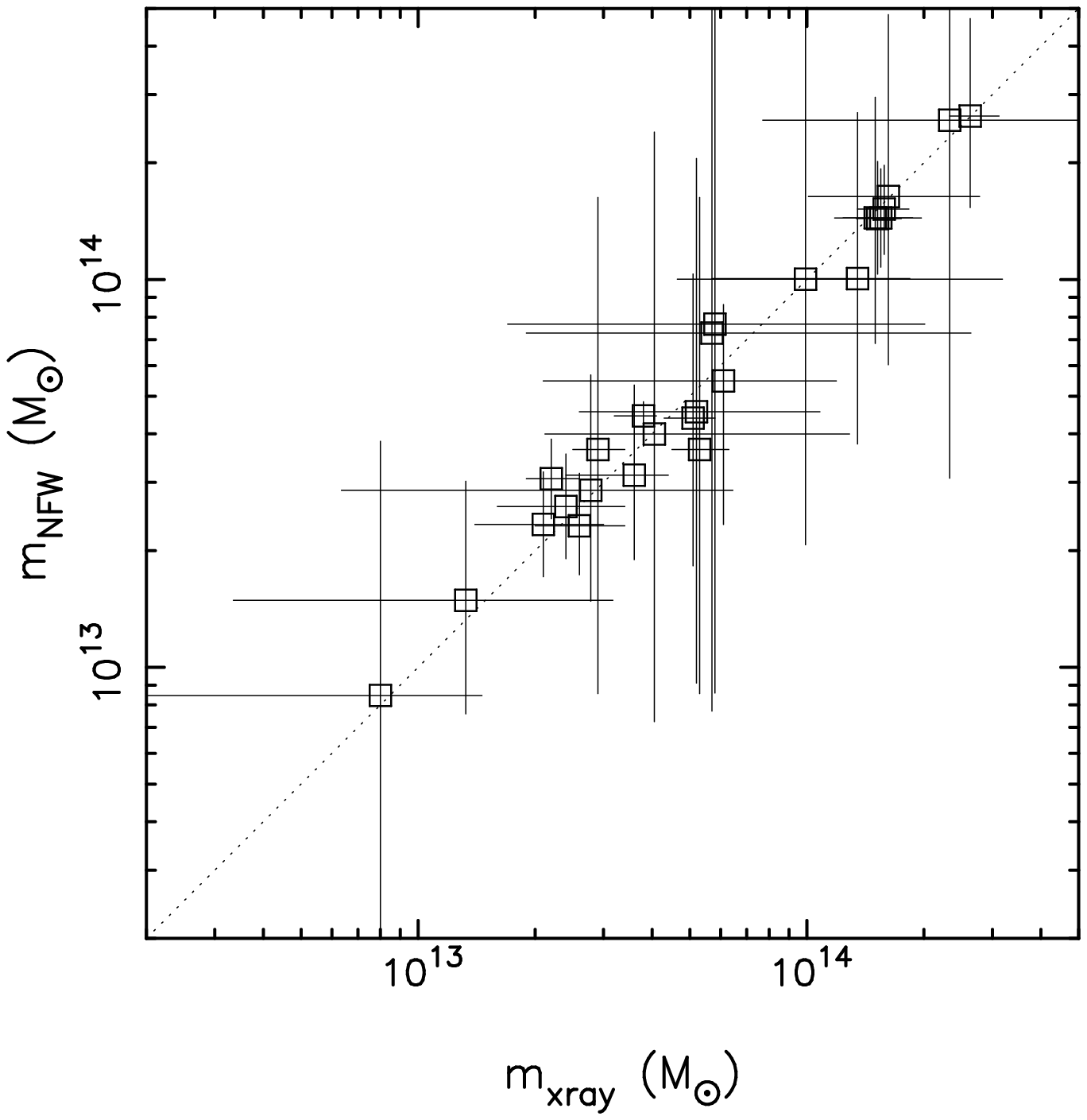,width=1.\textwidth,angle=270}}
\caption{The NFW profile derived cluster masses $m_{\rm NFW}$
within 26 arc-like images are plotted against 
the corresponding X-ray masses ($m_{\rm xray}$). Data are taken from Table 4.
}
\end{figure*}


\begin{thebibliography}{}

\bibitem{} Allen S.W., 1998, MNRAS, 296, 392
\bibitem{} Allen S.W., Fabian A.C., Edge A.C., Bautz M.W., Furuzawa A.,
	   Tawara Y., 1996, MNRAS, 283, 263	   
\bibitem{} Arnaud M., Evrard A.E., 1999, MNRAS, 305, 631
\bibitem{} Bartelmann M., Steinmetz M., 1996, MNRAS, 283, 431
\bibitem{} Carlberg R.G., Yee H.K.C., Ellingson E., 1997, ApJ, 478, 462
\bibitem{} Cole S., Lacey C., 1996, MNRAS, 281, 716
\bibitem{} Eke V.R., Navarro J.F., Frenk C.S., 1998, ApJ, 503, 569
\bibitem{} Elbaz D., Arnaud M., B\"ohringer H., 1995, A\&A, 293, 337
\bibitem{} Ettori S., Fabian A.C., 1999, MNRAS, 305, 834
\bibitem{} Gioia I.M., Shaya E.J., Le F\`evre O., Falco E.E., 
	   Luppino G.A., Hammer F., 1998, ApJ, 497, 573 
\bibitem{} Grossman S.A., Saha P., 1994, ApJ, 431, 74
\bibitem{} Hammer F., 1991, ApJ, 383, 66
\bibitem{} Lewis A.D., Ellingson E., Morris S.L., Carlberg R.G.,
	   1999, ApJ, 517, 587
\bibitem{} Loeb A.,  Mao S., 1994, ApJ, 435, L109
\bibitem{} Miralda-Escud\'e J., Babul A., 1995, ApJ, 449, 18
\bibitem{} Makino N., Asano K., 1999, 512, 9
\bibitem{} Makino N., Sasaki S., Suto Y., 1998, ApJ, 497, 55
\bibitem{} Mohr J.J., Mathiesen B., Evrard A.E., 1999, ApJ, 517, 627
\bibitem{} Molikawa K., Hattori M., Kneib J.-P., Yamashita K., 1999,
	   A\&A, 351, 418
\bibitem{} Navarro J.F., Frenk C.S., White S.D.M., 1995, MNRAS, 275, 720 (NFW)
\bibitem{} Navarro J.F., Frenk C.S., White S.D.M., 1997, ApJ, 490, 493
\bibitem{} Neumann D.M., Arnaud M., 1999, 348, 711
\bibitem{} Ota N., Mitsuda K., Fukazawa Y., 1998, ApJ, 495, 170 
\bibitem{} Rizza E., Burns J.O., Ledlow M.J., Owen F.N., Voges W.,
           Bliton M., 1998, MNRAS, 301, 328
\bibitem{} Schindler S., 1999, A\&A, 349, 435 
\bibitem{} Soucail G., Ota N., B\"ohringer H., Czoske O., Hattori M.,
           Mellier Y., 1999, A\&A, submitted (astro-ph/9911062)
\bibitem{} Vikhlinin A., Forman W., Jones C., 1999, ApJ, 525, 47
\bibitem{} Wu X.-P., 1994, ApJ, 436, L115
\bibitem{} Wu X.-P.,  Chiueh T., Fang L.-Z., Xue Y.-J., 1998, MNRAS, 301, 861
\bibitem{} Wu X.-P.,  Fang L.-Z., 1997, ApJ, 483, 62
\bibitem{} Wu X.-P.,  Hammer F., 1993, MNRAS, 262, 187
\bibitem{} Wu X.-P., Xue Y.-J., Fang L.-Z., 1999, ApJ, 524, 22 
\bibitem{} Wu X.-P., Xue Y.-J., 2000, ApJ, 529, L5 
\end{thebibliography}
\end{document}